\documentclass[aps,pre,twocolumn,groupedaddress,floatfix,longbibliography,superscriptaddress]{revtex4-1}

\usepackage{amsmath}
\usepackage{amssymb}
\usepackage{amsfonts}
\usepackage{graphicx}
\usepackage{bbold}
\usepackage{bm}
\usepackage{bm}
\usepackage{cancel}
\usepackage{times,float}
\usepackage{graphicx}
\usepackage[usenames,dvipsnames,svgnames]{xcolor}
\usepackage{hyperref}
\hypersetup{colorlinks=true, linkcolor=Blue, citecolor=Blue,urlcolor=Blue}
\usepackage{multirow}
\usepackage{ulem}
\usepackage{color,epsfig}
\usepackage{bm}
\usepackage{slashed}
\usepackage{hyperref}
\usepackage{subfigure}
\usepackage{feynmf}
\usepackage{array}
\usepackage{braket}

\newcommand{\bea}{\begin{eqnarray}}
\newcommand{\eea}{\end{eqnarray}}
\newcommand{\be}{\begin{equation}}
\newcommand{\ee}{\end{equation}}

\newcommand{\nn}{\nonumber}
\newcommand{\ii}{\mathrm{i}}

\newcommand{\Ham}{\hat{\mathcal{H}}}

\newcolumntype{P}[1]{>{\centering\arraybackslash}p{#1}}

\newcommand{\kz}{\widetilde{k}_z}
\newcommand{\R}{\mathcal{R}}
\newcommand{\lt}{\widetilde{\lambda}}
\newcommand{\zt}{\widetilde{z}}
\newcommand{\dt}{\widetilde{d}}
\usepackage[utf8]{inputenc}
\usepackage[T1]{fontenc}

\usepackage{feynmf}
\usepackage{braket}

\begin{document}
\title{Inversion symmetry breaking {in the probability} density by surface-bulk hybridization in topological insulators}
\author{Jorge David Casta\~no-Yepes}
\email{jcastano@uc.cl}
\affiliation{Facultad de F\'isica, Pontificia Universidad Cat\'olica de Chile, Vicu\~{n}a Mackenna 4860, Santiago, Chile}
\author{Enrique Mu\~noz}
\email{ejmunozt@uc.cl}
\affiliation{Facultad de F\'isica, Pontificia Universidad Cat\'olica de Chile, Vicu\~{n}a Mackenna 4860, Santiago, Chile}
\affiliation{Center for Nanotechnology and Advanced Materials CIEN-UC, Avenida Vicuña Mackenna 4860, Santiago, Chile}


\begin{abstract}
We analyze the probability density distribution in a topological insulator slab of finite thickness, where the bulk and surface states are allowed to hybridize. By using an effective continuum Hamiltonian approach as a theoretical framework, we analytically obtained the wave functions for each state near the $\Gamma$-point. Our results reveal that, under particular combinations of the hybridized bulk and surface states, the spatial symmetry of the electronic probability density with respect to the center of the slab can be spontaneously broken. This symmetry breaking arises as a combination of the parity of the solutions, their spin projection, and the material constants.
\end{abstract}

\maketitle
%
%


\section{Introduction}

Topological insulators (TIs) have emerged as fascinating materials with unique electronic properties that hold great promise for potential applications in electronic transport, spintronics~\cite{RevModPhys.82.3045} and quantum computation~\cite{Fu_2008}. Characterized by a large band-gap in the bulk, they possess topologically protected surface states with pseudo-relativistic (Dirac) dispersion that confer them non-trivial conducting properties~\cite{RevModPhys.82.3045}. The interplay between the bulk and surface states in TIs plays a pivotal role in shaping their behavior and functionality~\cite{moore2010birth}.

Previous studies have focused on the properties of surface states in topological insulators, providing valuable insights into their behavior. Extensive exploration has been carried out on confined electronic states and optical transitions in nanoparticles with different geometries~\cite{CASTANOYEPES2022105712,PhysRevB.86.235119,tian2013dual,Governale_2020}. From both theoretical and experimental perspectives, the spin properties of surface states have garnered attention in the field of the so-called quantum spin-Hall effect~\cite{PhysRevLett.101.246807,PhysRevB.81.115407}. Furthermore, there is evidence suggesting the existence of residual bulk carriers that couple with surface states through disorder~\cite{PhysRevB.90.245418,PhysRevB.85.121103,Mandal_2023,Yang2020}. On the other hand, it has been observed that charge carriers in TI nanostructures consist of both surface and bulk electronic states~\cite{PhysRevLett.106.196801,chiatti20162d,liao2017enhanced,PhysRevB.84.233101,zhao2013demonstration}. These studies point to the hybridization and interaction between topologically trivial and non-trivial states. The precise mechanism governing their coupling is not yet fully understood, but its comprehension is vital for elucidating the intricate nature of transport and photoemission experiments~\cite{PhysRevB.91.041401,10.21468/SciPostPhysCore.5.1.017,Singh_2017}. { Earlier studies have described the surface to bulk hybridization mechanism via a Fano model, with a phenomenological coupling constant~\cite{Hsu_014}, combined with ab-initio calculations in a slab geometry. Those results show that, despite the hybridization with the bulk state, the spin polarization of the surface state tends to be preserved by this mechanism~\cite{Hsu_014}.}

In this study, we investigate the hybridization between surface and bulk states in TIs, calculated from a paradigmatic continuous Hamiltonian model~\cite{Liu_10,Zhang_09} in the geometry of a flat slab of finite thickness $d$. {First principles calculations~\cite{Reid_2020} in this particular geometry point towards the relevance of the slab thickness on the size of the gap and for the existence of the surface state.}
Our focus lies in the observation of a significant asymmetry in the electronic density, resulting from the hybridization between bulk and surface states, and we highlight the role of parity and spin in this simple but rather surprising effect. Interestingly, this analytical result may provide an interpretation for the symmetry breaking effect observed in magnetotransport experiments in $\text{Bi}_2 \text{Se}_3$ nanoplates on $\text{Fe}_3 \text{O}_4$ substrates~\cite{Buchenau_2017}.

\section{The effective Hamiltonian for a Topological Insulator slab}

To analyze the possibility for hybridization between surface and bulk states in a topological insulator, along with its physical consequences, we shall start by considering a generic, low-energy continuum model obtained and reported in the literature~\cite{Liu_10,Zhang_09} by $(\mathbf{k}\cdot\mathbf{p})$-perturbation theory, that represents the band structure of a typical TI near the $\Gamma$-point~\cite{Liu_10,Zhang_09}:
\bea
\Ham&=&\left(\frac{\Delta_0}{2}+\frac{\gamma}{k_\Delta}\mathbf{k}^2\right)\hat{\tau}_z\otimes\hat{\sigma}_0\nn\\
&+&\gamma\left(k\hat{\tau}_x\otimes\hat{\sigma}_z+k_-\hat{\tau}_x\otimes\hat{\sigma}_++k_+\hat{\tau}_x\otimes\hat{\sigma}_-\right),
\label{Hamiltonian}
\eea
where the spin ($\hat{\sigma}$) and pseudo-spin ($\hat{\tau}$) degrees of freedom are incorporated. Here, $\mathbf{k}=(k_x,k_y,k_z)=-\ii\nabla$ represents the momentum operator, with $k_\pm=k_x\pm \ii k_y$ combinations of its components, while $\hat{\sigma}_\pm=(\hat{\sigma}_x\pm\ii\hat{\sigma}_y)/2$ are the ladder spin operators. As a reference, representative numerical values for the parameters $\Delta_0$, $\gamma$, and $k_{\Delta}$ for three different TI materials are presented in Table~\ref{tab:parameters}. 

\begin{table}[H]
\begin{ruledtabular}
\begin{tabular}{|l|c|c|c|c|c|c|}
 Material & $\Delta_0$ (eV) & $\gamma$ (eV Å) & $k_{\Delta}$ (Å$^{-1}$) & $R_0$ (Å) & $a$ (Å) & $c$ (Å) \\
\hline
 Bi$_2$ Se$_{3}$ & -0.338 & 2.2 & 0.13 & 1.3 & $4.140^{a}$ & $28.657^{a}$ \\
  &  &  &  &  & $4.143^{b}$ & $28.636^{b}$\\
 Bi$_2$ Te$_{3}$ & -0.3 & 2.4 & 0.026 & 0.81 & 4.400$^{a}$ & 30.096$^{a}$ \\
  &  &  &  &  & 4.386$^{b}$ & 30.497$^{b}$ \\
\end{tabular}
\end{ruledtabular}
\caption{\label{tab:parameters}Parameters for the effective Hamiltonian Eq.~\eqref{Hamiltonian} for different topological insulators after Refs.~\cite{Liu_10,Nechaev_16,Assaf_17}, {including lattice constants: $^a$DFT~\cite{Reid_2020}, $^{b}$Experiment~\cite{NAKAJIMA_63}.} }
\end{table}

In order to simplify the mathematical notation, and further algebraic manipulations, we define the ``mass" differential operator
\bea
\mathbb{M}(\mathbf{k})\equiv\frac{\Delta_0}{2}+\frac{\gamma}{k_\Delta}\mathbf{k}^2,
\label{eq:mop}
\eea
so that the Hamiltonian Eq.~\eqref{Hamiltonian} can be written in the alternative and more compact form
\bea
\Ham=\mathbb{M}(\mathbf{k})\hat{\beta}+\gamma\left(\boldsymbol{\alpha}\cdot\mathbf{k}\right),
\label{eq:Ham2}
\eea
where we defined the Dirac $4\times4$ matrices in the standard representation
\bea
\boldsymbol{\alpha}=\begin{pmatrix}
0 & \boldsymbol{\hat{\sigma}}\\
\boldsymbol{\hat{\sigma}} & 0\end{pmatrix},~~~\hat{\beta}=\begin{pmatrix}
\mathbb{I} & 0\\
0& -\mathbb{I}\end{pmatrix},
\eea
with $\boldsymbol{\hat{\sigma}} = \left( \hat{\sigma}_x,\hat{\sigma}_y, \hat{\sigma}_z\right)$.
The representation Eq.~\eqref{eq:Ham2} also provides the conceptual advantage to possess the explicit form of a pseudo-relativistic Dirac Hamiltonian with a momentum-dependent "mass" operator as defined by Eq.~\eqref{eq:mop}. One can thus make the appropriate analogies to interpret the physical meaning of the parameters involved, where $\Delta_0$ clearly represents the "mass" that defines the finite gap of the bulk spectrum at the $\Gamma$-point, while $\gamma = \hbar v_F$ is proportional to the Fermi velocity, and $k_{\Delta}$ controls the curvature of the small quadratic correction to the linear dispersion relation at such point.
We remark that an effective continuum model, such as Eq.~\eqref{Hamiltonian} or its equivalent form Eq.~\eqref{eq:Ham2}, constitutes a fair representation of the physics at length-scales larger than the lattice constant of the material. As shown in Table~\ref{tab:parameters}, the lattice constants for the two metal trichalcogenides Bi$_{2}$Se$_{3}$ and Bi$_{2}$Te$_{3}$ are on the order of $\sim 4$\r{A}. Based on these reference values, a continuum model representation is fair for confinement length scales on the order of $d\gtrsim 5 R_0$, where $R_0\equiv 2\gamma/|\Delta_0|$ is a material-dependent parameter whose pseudo-relativistic interpretation in the context of the Dirac Hamiltonian is the "Compton wavelength", i.e. a characteristic length-scale for the support of the spinor eigenstates. 

We assume that our system is a planar slab of a topological insulator with thickness $d$, whose Hamiltonian in dimensionless form can be written as
\bea
\widetilde{\mathcal{H}}=\left(1+\frac{s}{\mathcal{R}}\widetilde{\mathbf{k}}^2\right)\hat{\beta}+s(\mathbf{\widetilde{k}}\cdot\hat{\boldsymbol{\alpha}}),
\eea
where $\widetilde{\mathcal{H}}\equiv2\Ham/\Delta_0$,  $\widetilde{k}\equiv R_0\hat{k}$, $\mathcal{R}\equiv k_\Delta R_0$, and $s=\text{sign}(\Delta_0)$.
In this dimensionless representation of the parameters, the corresponding eigenvalue problem is stated by the equation
\bea
\widetilde{\mathcal{H}}\hat{\Psi}(\mathbf{r}_\perp,z)=\epsilon\hat{\Psi}(\mathbf{r}_\perp,z).
\label{eq_eig_gen}
\eea

{From the geometry of our system, the translational invariance on the plane $(x,y) = \mathbf{r}_\perp$ normal to the interfaces at $z = 0$ and $z=d$, respectively, allows us for separation of variables in the eigenfunction
\bea
\hat{\Psi}(\mathbf{r}_\perp,z)=e^{\ii\mathbf{k}_\perp\cdot\mathbf{r}_\perp}\Psi(z),
\label{eq_factor_spinor}
\eea
where $\mathbf{k}_\perp=(k_x,k_y)$, and $\Psi(z)$ is a four-component spinor. This separation of variables leads to the secular equation
{\small \bea
\left[1+\frac{s}{\mathcal{R}}\left(\widetilde{\mathbf{k}}_\perp^2+k_z^2\right)\hat{\beta}+s(\mathbf{\widetilde{k}_\perp}\cdot\hat{\boldsymbol{\alpha}}_\perp+\widetilde{k}_z\hat{\alpha}_z)\right]\Psi(z)=\epsilon\Psi(z),
\eea}
where we defined the dimensionless energy eigenvalues
\bea
\epsilon\equiv\frac{E}{\Delta_0/2}.
\label{Eq.Energydef}
\eea

The 4-component spinor eigenstates can be expressed in terms of its bi-spinor components $\chi$ and $\eta$, i.e. $\Psi(z)= e^{\ii k_z z}(\eta,\chi)^T$, such that the eigenvalue problem Eq.~\eqref{eq_eig_gen} spans a system of coupled algebraic equations
\bea
\left[1+\frac{s}{\mathcal{R}}\left(\widetilde{\mathbf{k}}_\perp^2+k_z^2\right)-\epsilon\right]\eta+s\left(\widetilde{\mathbf{k}}_\perp\cdot\hat{\boldsymbol{\sigma}}_\perp + \widetilde{k}_z\hat{\sigma}_z\right)\chi&=&0\nn\\
\left[1+\frac{s}{\mathcal{R}}\left(\widetilde{\mathbf{k}}_\perp^2+k_z^2\right)+\epsilon\right]\chi-s\left(\widetilde{\mathbf{k}}_\perp\cdot\hat{\boldsymbol{\sigma}}_\perp + \widetilde{k}_z\hat{\sigma}_z\right)\eta&=&0.\nn\\
\label{eq_Dirac_system}
\eea

The latter system of equations implies that the bi-spinors $\eta$ and $\chi$ are not independent. Indeed, for a generic $(\tilde{\mathbf{k}}_{\perp},\tilde{k}_z)$, it is straightforward to obtain from the second equation in the system Eq.~\eqref{eq_Dirac_system}
\bea
\chi=\frac{s\left(\kz\hat{\sigma}_z + \widetilde{\mathbf{k}}_\perp\cdot\hat{\boldsymbol{\sigma}}_\perp\right)}{1+\frac{s}{\R}\left(\kz^2 + \widetilde{\mathbf{k}}_\perp^2 \right)+\epsilon}\eta,
\label{chi_termsof_phi_1}
\eea
and the corresponding energy eigenvalues are functions of $\tilde{k}_z^2 + \widetilde{\mathbf{k}}_\perp^2$, as follows
\bea
\epsilon^2=\kz^2 + \widetilde{\mathbf{k}}_\perp^2 +\left[1+\frac{s}{\R}\left(\kz^2 + \widetilde{\mathbf{k}}_\perp^2\right)\right]^2.
\label{eq:energy_general_3D}
\eea

Clearly then, the contribution to the energy eigenvalues arising from the transverse degrees of freedom is a monotonically increasing function of $\mathbf{\widetilde{k}_\perp}^2$. Therefore, since we are exploring the consequences of spontaneous surface-bulk hybridization, we shall be interested in the lowest energy eigenvalues, and hence we choose $\mathbf{\widetilde{k}_\perp} = 0$, with the corresponding consequence on the transverse plane-wave phase of the spinor function in Eq.~\eqref{eq_factor_spinor}. Therefore, by setting $\mathbf{\widetilde{k}_\perp} = 0$ in Eq.~\eqref{eq:energy_general_3D}, the general energy dispersion relation that we shall use in what follows is}
\bea
\epsilon^2=\kz^2+\left(1+\frac{s}{\R}\kz^2\right)^2.
\label{eq:energy_general}
\eea

\section{The surface states}

\begin{figure}[h!]
    \centering
    \includegraphics[width=0.45\textwidth]{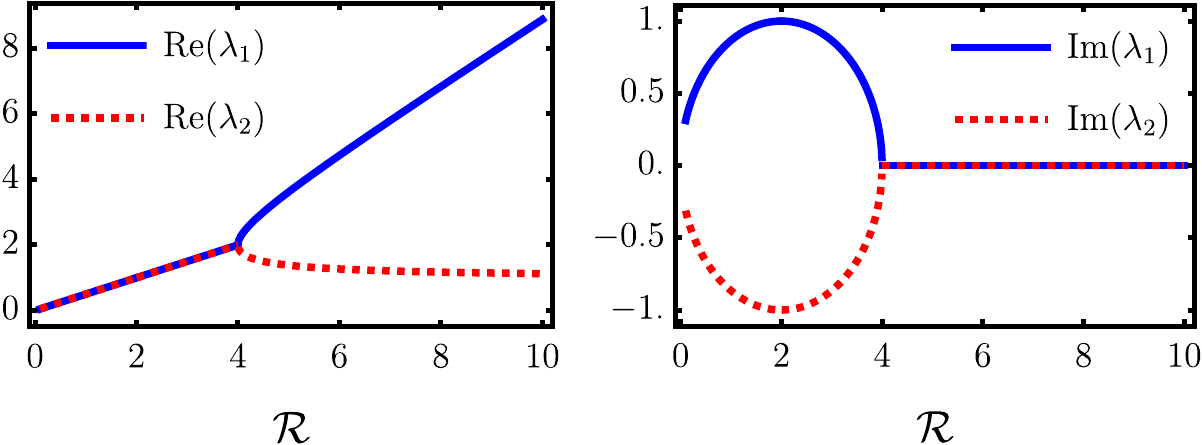}
    \caption{{Real and imaginary parts of $\lambda_1$, and $\lambda_2$ defined in Eq.~\eqref{eq_lambdas} for energies $\epsilon\simeq0$ corresponding to surface eigenstates.}}
    \label{fig:lambdas}
\end{figure}

The surface states are characterized by an exponential localization near the boundaries $z=0$ and $z=d$, respectively. This behaviour
is captured by choosing $\tilde{k}_z=\ii\tilde{\lambda}$ a complex number. Under this ansatz, Eq.~\eqref{eq:energy_general} admits four possible solutions for  $\tilde{k}_z$, given by $\pm\tilde{\lambda}_1$, and $\pm\tilde{\lambda}_2$, so that:
\bea
\lt^2_\alpha=\frac{\R^2}{2}\left[1+\frac{2s}{\R}+(-1)^{1+\alpha}\sqrt{1+\frac{4}{\R}\left(s+\frac{\epsilon^2}{\R}\right)}\right],
\label{eq_lambdas}
\eea
where we used the dimensionless form $\lt=R_0\lambda$. {Figure~\ref{fig:lambdas} shows how each $\lambda$ changes as a function of the material-dependent parameter $\R$. In particular, as can be trivially verified from Eq.~\eqref{eq_lambdas}, for surface states ($\epsilon \simeq 0$) the imaginary part of $\lambda_{1,2}$ vanishes when $\R > 4$, while it remains present for smaller values.} 

Then, by solving the linear system Eq.~\eqref{eq_Dirac_system}, we obtain four independent solutions (for $\alpha = 1,2$ and $\zt=z/R_0$)
\bea
\varphi_\pm^{(\alpha,\uparrow)}(\zt,\epsilon)&=&e^{\pm\lt_\alpha\zt}\left(
    1,0,\pm\ii A_\alpha,0\right)^T,\nn\\
\varphi_\pm^{(\alpha,\downarrow)}(\zt,\epsilon)&=&e^{\pm\lt_\alpha\zt}\left(0,1,0,\mp\ii A_\alpha
\right)^T,
\label{eq:spin_components_surface}
\eea 
where we defined the parameters
\bea
A_\alpha=\frac{\lt_\alpha}{1-\frac{s}{\R}\lt^2_\alpha+\epsilon}.
\eea

{The exponential decay length in Eq.~\eqref{eq:spin_components_surface} is thus a non-linear function of the energy eigenvalues after Eq.~\eqref{eq_lambdas}. In particular, for $\R \gg 1$ after Eq.~\eqref{eq_lambdas} it is clear that $\tilde{\lambda}_{\alpha}\sim \R/\sqrt{2}$, i.e. the exponential decay of the surface states from each side of the slab is approximately determined by the magnitude of this parameter. Concerning the materials we present in Table~\ref{tab:parameters}, realistic values are $\mathcal{R} = 0.169$ for ${\rm{Bi}}_2 {\rm{Se}}_3$ and $\mathcal{R} = 0.021$ for ${\rm{Bi}}_s {\rm{Te}}_3$, respectively, and hence the imaginary part of the $\lambda_{1,2}$ does not vanish in both cases. Moreover, based on these material specific examples, we shall focus on the analysis for representative values $\mathcal{R} \le 1$.}

We remark that these independent solutions are trivially orthogonal for opposite spin components, i.e. (for $\beta,\,\beta'=\pm$, $\alpha,\,\alpha' = 1,2$)
\be
\langle \alpha \uparrow;\beta|\alpha'\downarrow;\beta'\rangle = \int_0^{\tilde{d}}d\tilde{z}\left[\varphi^{(\alpha,\uparrow)}_{\beta}(\tilde{z})\right]^{\dagger}\varphi^{(\alpha',\downarrow)}_{\beta'}(\tilde{z})=
0,
\label{eq_ortho}
\ee
while those with parallel spin components satisfy the inner product result
\bea
\langle \alpha \downarrow;\beta|\alpha'\downarrow;\beta'\rangle &=&\langle \alpha \uparrow;\beta|\alpha'\uparrow;\beta'\rangle\nn\\ 
&=& \tilde{d}\left( 1 + \beta\beta'A_{\alpha}A_{\alpha'}  \right)
e^{\left(\beta\tilde{\lambda}_{\alpha}+\beta'\tilde{\lambda}_{\alpha'}\right)\tilde{d}/2}\nn\\
&&\times\frac{\sinh\left[ \left(\beta\tilde{\lambda}_{\alpha}+\beta'\tilde{\lambda}_{\alpha'}\right)\tilde{d}/2 \right]}{\left(\beta\tilde{\lambda}_{\alpha}+\beta'\tilde{\lambda}_{\alpha'}\right)\tilde{d}/2}.
\label{eq_ipsurf}
\eea

\begin{figure*}[h!]
    \centering
    \includegraphics[width=1\textwidth]{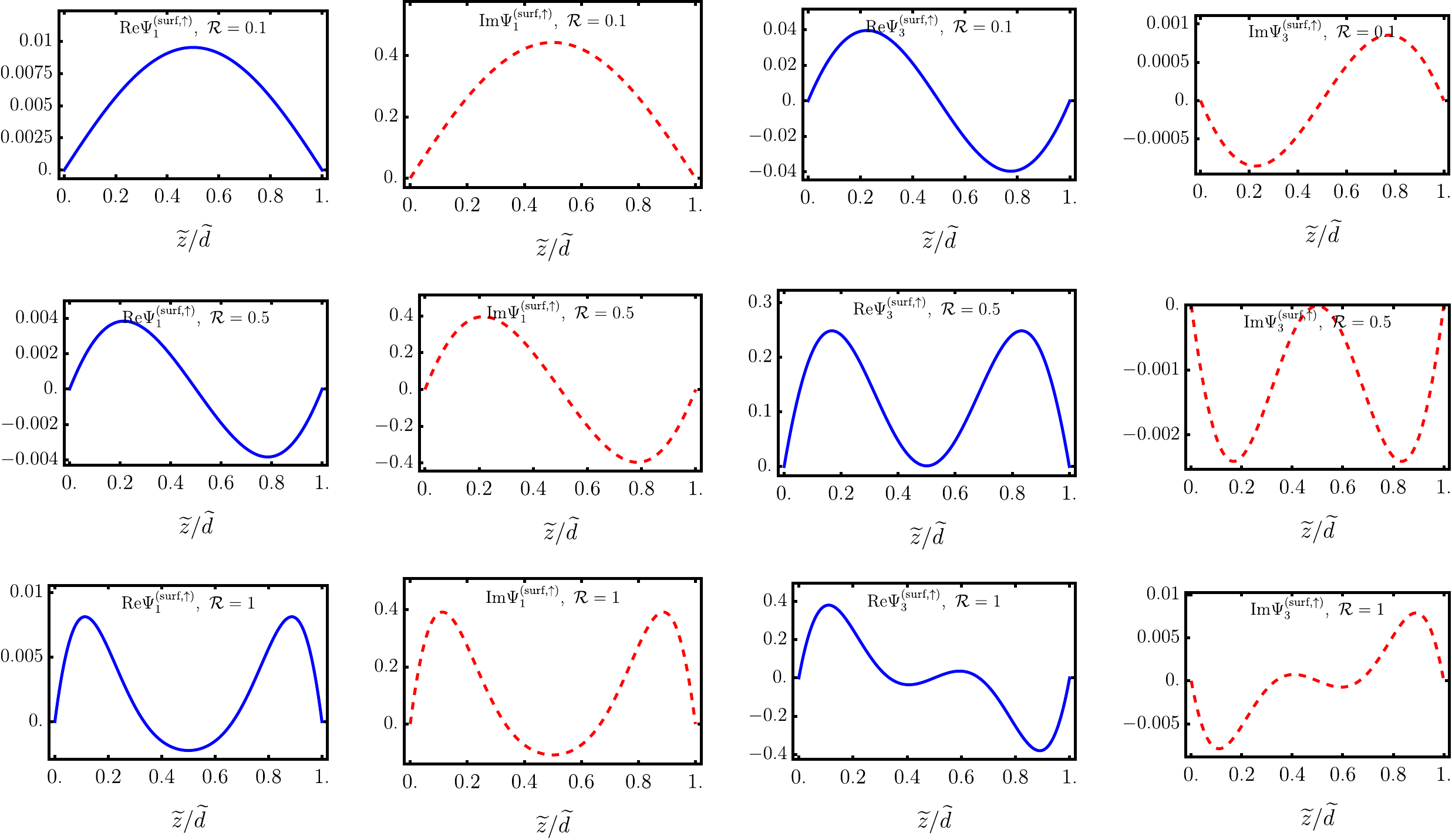}
    \caption{{Real and imaginary parts of the surface eigenstate $\Psi_\text{surf}^{\uparrow}=\left[\Psi_1^{(\text{surf},\uparrow)},0,\Psi_3^{(\text{surf},\uparrow)},0\right]^T$, for different values of $\R$.}}
    \label{fig:Parity_components}
\end{figure*}

\begin{figure*}[h!]
    \centering
    \includegraphics[width=1\textwidth]{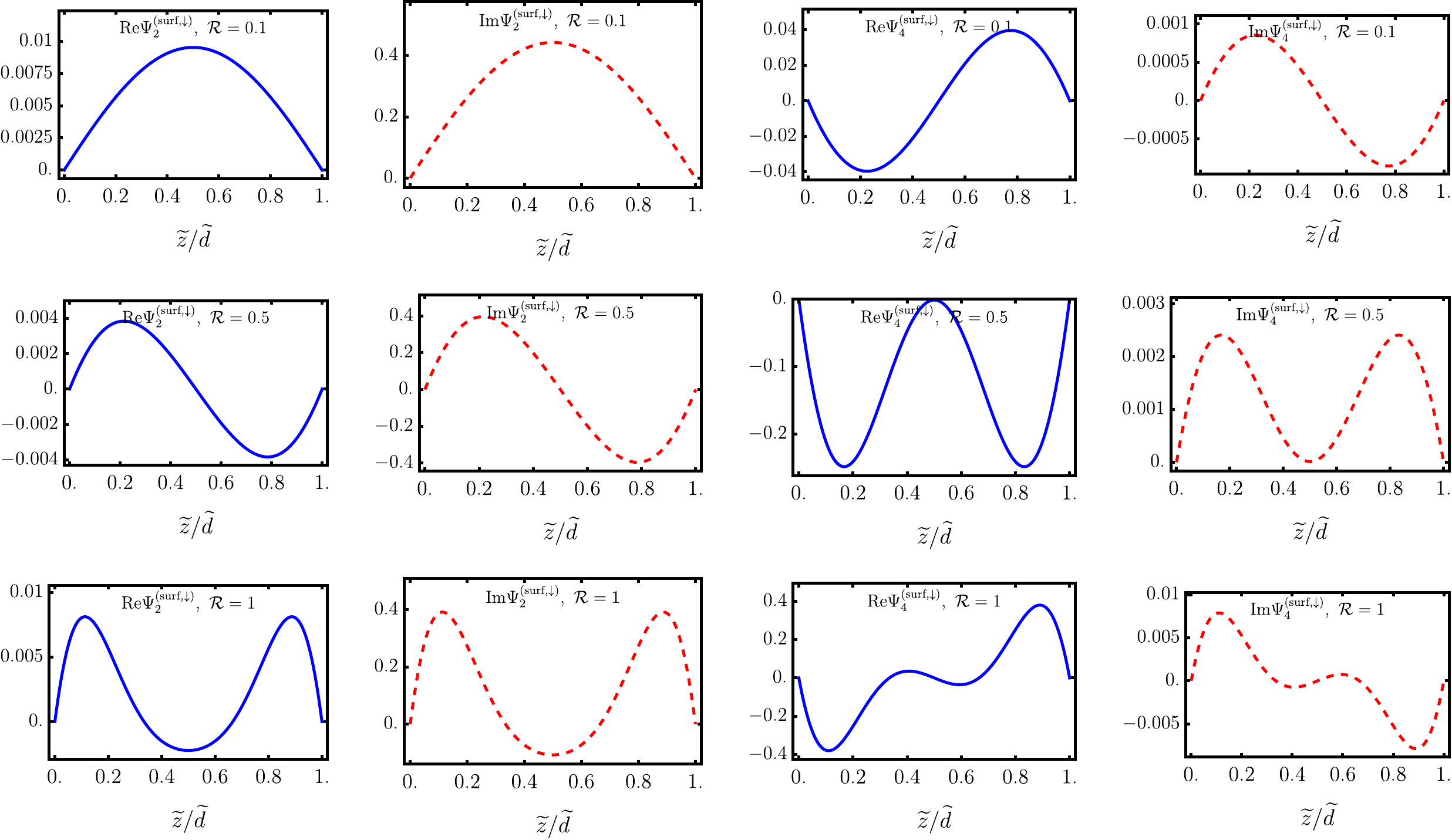}
    \caption{{Real and imaginary parts of the surface eigenstate $\Psi_\text{surf}^{\downarrow}=\left[0,\Psi_2^{(\text{surf},\downarrow)},0,\Psi_4^{(\text{surf},\downarrow)}\right]^T$, for different values of $\R$.}}
    \label{fig:Parity_components_down}
\end{figure*}

Therefore, it is natural to define independent surface states for each fixed spin projection $\sigma = \uparrow$ or $\sigma = \downarrow$ as follows
\bea
\Psi_\text{surface}^{(\sigma)}(\zt)=\sum_{\alpha=1,2}\sum_{\beta=\pm}a_\beta^{(\alpha)} \varphi_\beta^{(\alpha,\sigma)}(\zt),
\label{eq:linearcombination}
\eea
where $a_\beta^{(\alpha)}$ represent arbitrary coefficients in the general linear combination. Those coefficients are further defined upon imposing, for each spin projection state $\sigma = \uparrow$ or $\sigma = \downarrow$, the boundary conditions at each surface of the slab, 
\bea
\Psi_\text{surface}^{(\sigma)}(\zt=0)=\Psi_\text{surface}^{(\sigma)}(\zt=d/R_0)=0.
\label{BC_psi_up_z}
\label{BoundaryConditions}
\eea

Those lead to the secular linear system
\bea
\mathbb{A}~\boldsymbol{a}=0,
\label{systemBsz}
\eea
where $\boldsymbol{a}\equiv\left(a_+^{(1)},a_-^{(1)},a_+^{(2)},a_-^{(2)}\right)^T$ is the vector of coefficients, and the matrix $\mathbb{A}$ is defined as (for $\dt\equiv d/R_0$)
\bea
\mathbb{A}\equiv\begin{pmatrix}
1 & 1 & 1 & 1 \\
\ii A_1  & -\ii A_1  & \ii A_2 & -\ii A_2 \\
e^{\lt_1 \dt} & e^{-\lt_1 \dt} & e^{\lt_2 \dt} & e^{-\lt_2 \dt} \\
\ii A_1e^{\lt_1 \dt} & -\ii A_1 e^{-\lt_1 \dt} & \ii A_2e^{\lt_2 \dt} & -\ii A_2e^{-\lt_2 \dt}
\end{pmatrix}.
\eea

The existence of non-trivial solutions for the coefficients $\boldsymbol{a}$ that satisfy the boundary conditions Eq.~\eqref{BC_psi_up_z} is granted provided $\det\mathbb{A}=0$. This equation can be expressed in a closed algebraic form as
\bea
&&\frac{\lt_1}{\lt_2}\frac{1+\epsilon -\lt_2^2/\R}{1+\epsilon - \lt_1^2/\R}v(\lt_1,\lt_2 )= u(\lt_1,\lt_2)\nn\\
&&-\sqrt{u^2(\lt_1,\lt_2) - v^2(\lt_1,\lt_2)},
\label{eq_deter}
\eea
with the definitions {\small $u(\lt_1,\lt_2)\equiv\cosh( \dt~\lt_1)\cosh( \dt~\lt_2) -1$} and {\small$v(\lt_1,\lt_2)\equiv\sinh( \dt~\lt_1 )\sinh( \dt~\lt_2)$}, respectively.

\begin{widetext}
Furthermore, from the first three rows of Eq.~\eqref{systemBsz} we obtain
   \bea
\bar{a}_{-}^{(1)}&\equiv& a_-^{(1)}/a_+^{(1)}=\frac{\left(e^{2\dt~\lt_2}-1\right)e^{\dt~\lt_1}A_1+\left(1+e^{2\dt~\lt_2}-2e^{\dt(\lt_1+\lt_2)}\right)e^{\dt~\lt_1}A_2}{e^{\dt~\lt_1}\left(e^{2\dt~\lt_2}-1\right)A_1-\left(e^{\dt~\lt_1}-2e^{\dt~\lt_2}+e^{\dt(\lt_1+2\lt_2)}\right)A_2}\nonumber\\
\bar{a}_{+}^{(2)}&\equiv& a_+^{(2)}/a_+^{(1)}=\frac{\left(e^{2\dt~\lt_1}-1\right)e^{\dt~\lt_2}A_2-\left(e^{\dt~\lt_2}-2e^{\dt~\lt_1}+e^{\dt(2\lt_1+\lt_2)}\right)A_1}{e^{\dt~\lt_1}\left(e^{2\dt~\lt_2}-1\right)A_1-\left(e^{\dt~\lt_1}-2e^{\dt~\lt_2}+e^{\dt(\lt_1+2\lt_2)}\right)A_2}\nonumber\\
\bar{a}_{-}^{(2)}&\equiv& a_-^{(2)}/a_+^{(1)}=\frac{\left(e^{2\dt~\lt_1}-1\right)e^{\dt~\lt_1}A_2+\left(1+e^{\dt~\lt_1}-2e^{\dt(\lt_1+\lt_2)}\right)e^{\dt~\lt_1}A_1}{e^{\dt~\lt_1}\left(e^{2\dt~\lt_2}-1\right)A_1-\left(e^{\dt~\lt_1}-2e^{\dt~\lt_2}+e^{\dt(\lt_1+2\lt_2)}\right)A_2}.
\label{eq_atildes}
\eea 
\end{widetext}

Note that the constants in Eq.~\eqref{eq:linearcombination} are the same for both spin projections, 
so we can finally write
\bea
\Psi_\text{surface}^{(\uparrow)}(\zt,\epsilon)&=& C_{\uparrow}\sum_{\alpha=1,2}\sum_{\beta=\pm}\bar{a}_\beta^{(\alpha)} \varphi_\beta^{(\alpha,\uparrow)}(\zt)\nonumber\\
\Psi_\text{surface}^{(\downarrow)}(\zt.\epsilon)&=& C_{\downarrow}\sum_{\alpha=1,2}\sum_{\beta=\pm}\bar{a}_\beta^{(\alpha)} \varphi_\beta^{(\alpha,\downarrow)}(\zt).
\eea

Here, we defined $\bar{a}_{+}^{(1)} = 1$, and the coefficients $C_{\uparrow}$, $C_{\downarrow}$ are determined by the normalization conditions as follows
{\small\bea
|C_{\uparrow}| &=& \left[\sum_{\alpha = 1,2}\sum_{\alpha'=\pm}\sum_{\beta=1,2}\sum_{\beta'=1,2}\bar{a}_{\beta}^{(\alpha)}\bar{a}_{\beta'}^{(\alpha')}\langle \alpha \uparrow;\beta| \alpha'\uparrow;\beta'\rangle\right]^{-1/2}\nn\\
|C_{\downarrow}| &=& \left[\sum_{\alpha = 1,2}\sum_{\alpha'=\pm}\sum_{\beta=1,2}\sum_{\beta'=1,2}\bar{a}_{\beta}^{(\alpha)}\bar{a}_{\beta'}^{(\alpha')}\langle \alpha \downarrow;\beta| \alpha'\downarrow;\beta'\rangle\right]^{-1/2},
\eea}
where the analytical expression for the inner product coefficients are presented in Eq.~\eqref{eq_ipsurf}. Moreover, from Eq.~\eqref{eq_ortho} it is clear that surface states with opposite spin components are mutually orthogonal
\bea
&&\langle \uparrow;\text{surface},\epsilon|\downarrow;\text{surface},\epsilon\rangle\nn\\
&=& \int_{0}^{\tilde{d}}d\tilde{z}\left[\Psi_\text{surface}^{(\uparrow)}(\tilde{z},\epsilon)\right]^{\dagger}\Psi_\text{surface}^{(\downarrow)}(\tilde{z},\epsilon) = 0.
\eea

The system of equations Eq.~\eqref{eq_lambdas} and Eq.~\eqref{eq_deter} must be solved in favor of $\epsilon$. By inspection, it is evident that a solution is found when
\bea
\dt~\lt_2=0~~\text{and}~~\epsilon=-1
\eea
so that, this condition implies, after Eq.~\eqref{eq_lambdas}
\bea
\lt_1=\R\left(1+\frac{2s}{\R}\right).
\eea

On the other hand, nontrivial solutions corresponding to {\it true} surface states can exist when $\epsilon \simeq 0$, and those can only be found numerically as a function of the system thickness $d$.

{An important property of these surface states, determined from the numerical solutions of the algebraic system of equations detailed in this section, is the well-defined parity of their components with respect to the center of the slab at $z = d/2$, i.e.
\be
\Psi^{(\text{surface},\sigma)}_j(d/2 -\zeta,\epsilon) = (-1)^P\Psi^{(\text{surface},\sigma)}_j(d/2 +\zeta,\epsilon)
\label{eq_surface_parity}
\ee
where $j$ labels the component ($j=1,\ldots,4$), and $\zeta \in [-d/2,d/2]$ represents the distance from the center of the slab. Here, the parity index $P$ is {\it{odd}} ($P=1$) or {\it{even}} ($P=2$), depending on the spin polarization of the surface state and the magnitude of $\mathcal{R}$. As shown in Fig.~\ref{fig:Parity_components}, for the $\sigma = \uparrow$ spin polarization, the only non-vanishing components are $j=1,3$. When $\mathcal{R} = 0.1$ and $\R = 1$, the component $j=1$ has even parity $P=2$, while the $j=3$ component possesses odd parity $P=1$. In contrast, we observe that the parities are reversed when $\R = 0.5$.

A similar behavior is observed in Fig.~\ref{fig:Parity_components_down} for the $\sigma = \downarrow$ spin polarization, when in this case it is the $j=2$ and $j=4$ components which are non-zero and alternate their well defined but opposite parities as a function of $\R$.  

Nevertheless, despite both components always have opposite parities, the corresponding probability density $\rho_{{\rm{surface}}}^{(\sigma)}(z) = \Psi_{{\rm{surface}}}^{\sigma\dagger}(z)\Psi_{{\rm{surface}}}^{\sigma}(z)$ for either spin polarization $\sigma$ remains symmetric with respect to the center of the slab}
\be
\rho_{{\rm{surface}}}^{(\sigma)}(d/2 - \zeta) =
\rho_{{\rm{surface}}}^{(\sigma)}(d/2 + \zeta).
\ee

\section{The bulk eigenstates}

The bulk states are obtained from the same system of equations Eq.~\eqref{eq_Dirac_system}, but with the $z$-component of momentum $\tilde{k}_z$ a real number. From Eq.~\eqref{chi_termsof_phi_1}, we have that the general solution is given by the linear combination of spinors
\bea
\Phi^\pm_{\text{bulk}}(z)&=&\begin{pmatrix}
\eta_\text{bulk}^\pm\\
\\
\chi_\text{bulk}^\pm
\end{pmatrix}\nn\\
&=&ue^{\pm\ii \tilde{k}_z \zt}\begin{pmatrix}
    1\\
    0\\
    B_z\\
    0
\end{pmatrix}+ve^{\pm\ii \tilde{k}_z \zt}\begin{pmatrix}
    0\\
    1\\
    0\\
    -B_z
\end{pmatrix},
\label{eq:Psi_bluk}
\eea
with
\bea
B_z=\frac{s \tilde{k}_z}{1+\frac{s}{\R}\tilde{k}_z^2+\epsilon_\text{bulk}}.
\label{eq_Bz}
\eea

Now, by applying the confinement boundary conditions
\bea
\phi_{\text{bulk}}(\zt = 0) = \phi_{\text{bulk}}(\zt = \tilde{d}) = 0,
\eea
we conclude that the appropriate linear combination (modulo a normalization constant) is 
\be
\phi_{\text{bulk}}(\zt) = \Phi^{+}_{\text{bulk}}(\zt)-\Phi^{-}_{\text{bulk}}(\zt) \propto \sin(\tilde{k}_z \zt), 
\ee
provided
\bea
\sin(\tilde{k}_z\tilde{d}) = 0,
\eea
which leads to the quantization condition for $\tilde{k}_z$
\bea
\kz=\frac{n\pi}{\dt},~\text{for}~n=\pm1,\pm2,\ldots
\label{eq_kq}
\eea

Therefore, from Eq.~\eqref{eq:energy_general} the quantized
energy eigenvalues due to spatial confinement in the region $0\le z \le d$ are given by
\bea
\epsilon_\text{bulk}^2=\frac{n^2\pi^2}{\dt^2}+\left(1+\frac{s}{\R}\frac{n^2\pi^2}{\dt^2}\right)^2.
\eea

The corresponding normalized bulk eigenstates for each spin projection are
\bea
\phi_\text{bulk}^{(n,\uparrow)}(\zt)&=&\sqrt{\frac{2}{(1+B_n^2)\tilde{d}}}\sin\left(\frac{ n\pi \zt}{\dt}\right)\left(1,0,B_n,0
\right)^T,\nonumber\\
\phi_\text{bulk}^{(n,\downarrow)}(\zt)&=&\sqrt{\frac{2}{(1+B_n^2)\tilde{d}}}\sin\left(\frac{ n\pi \zt}{\dt}\right)\left(0,1,0,-B_n\right)^T,\nn\\
\eea
with $\dt\equiv d/R_0$, 
and the coefficients $B_n$ defined as
\be
B_n = B_z\left(\tilde{k}_z = \frac{n\pi}{\dt}\right )=\frac{s n\pi/\dt}{1+\frac{s}{\R}\left(\frac{n\pi}{\dt}\right)^2+\epsilon_\text{bulk}}.
\ee

The normalization constants are defined such that (for $\sigma = \uparrow,\downarrow$)
\bea
\int_0^{\tilde{d}}d \zt \left[ \phi_{\text{bulk}}^{n,\sigma}(\zt) \right]^{\dagger}\phi_{\text{bulk}}^{n,\sigma}(\zt) = 1.
\eea

A noteworthy property of these spin projection eigenstates is their even/odd parity with respect to the center of the slab $z = d/2$, since
for $\zeta \in [-d/2,d/2]$ the distance {\it{from the center}} of the slab, we have
\be
\sin\left(\frac{n\pi}{d}\left( d/2 - \zeta \right)  \right) = (-1)^{n+1}\sin\left(\frac{n\pi}{d}\left( d/2 + \zeta \right)  \right).  
\ee

Therefore, the same even/odd parity is inherited by the spinor functions
\be
\phi_{\text{bulk}}^{n,\sigma}\left(d/2 - \zeta \right) = (-1)^{n+1}\phi_{\text{bulk}}^{n,\sigma}\left(d/2 + \zeta \right).
\label{eq_bulk_parity}
\ee

\section{Surface - bulk hybridization and the breaking of inversion symmetry}

\begin{figure*}
    \centering
    \includegraphics[width=1\textwidth]{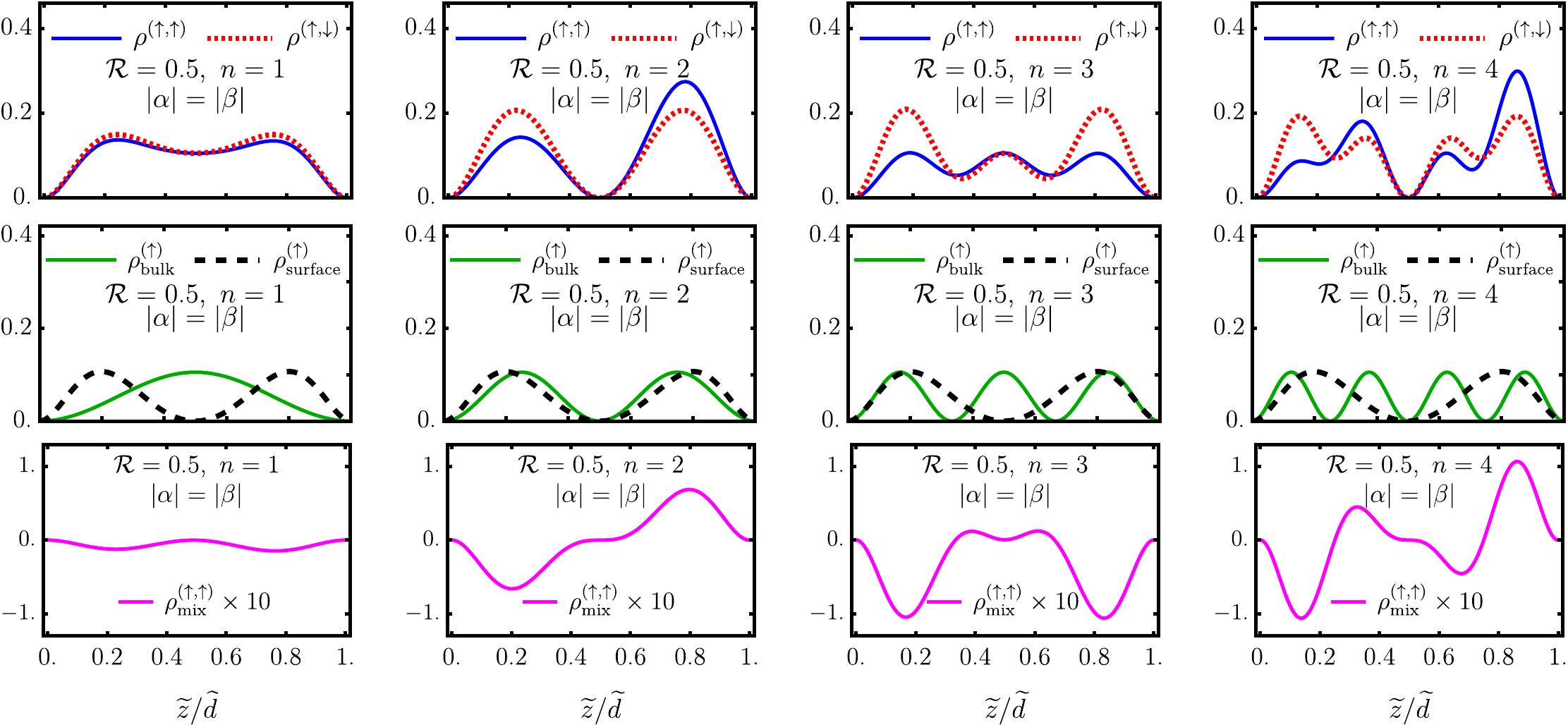}
    \caption{{Electronic density distribution $\rho_{n,\epsilon}^{\sigma,\sigma'}(\zt)$ and its independent surface $\rho_{\text{surface}}$, bulk $\rho_{\text{bulk}}$, and interference $\rho_{\text{mix}}$ components, respectively, as defined in Eq.~\eqref{eq_rhohybrid} for the coupling between $(\sigma,\sigma')=(\uparrow,\uparrow)$ and $(\uparrow,\downarrow)$. Note that if $\sigma\neq\sigma'\to\rho_\text{mix}^{(\sigma,\sigma')}=0$.}}
    \label{fig:Psi2VsnR1mix}
\end{figure*}

\begin{figure}
    \centering
    \includegraphics[width=0.4\textwidth]{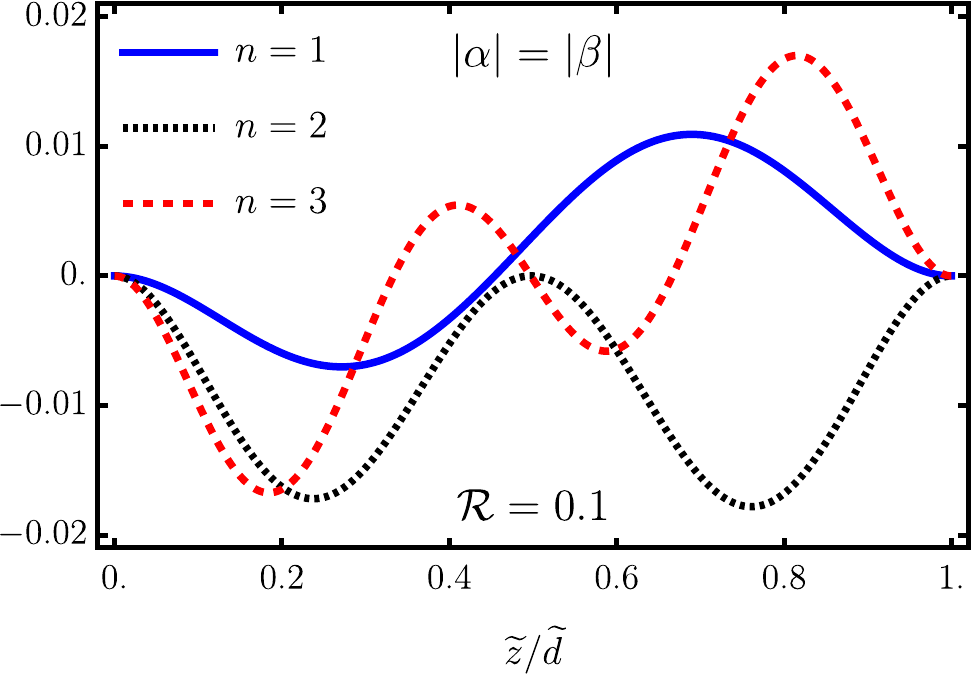}
    \caption{{Behavior of $\rho_\text{mix}^{(\uparrow,\uparrow)}$ as a function of $\R$, and $n$.}}
    \label{fig:Psi_mix2}
\end{figure}

As discussed in Eq.~\eqref{eq_surface_parity}, depending on the magnitude of $\mathcal{R}$, the spinor components of the surface states possess well defined but opposite parities $P$ with respect to a mirror plane located at the center of the slab $z = d/2$. On the other hand, the bulk states exhibit a global parity determined by the number of nodes within the finite domain $z\in[0,d]$ imposed by the boundaries of the slab, which depends on the quantum number $n \ge 1$, as stated in Eq.~\eqref{eq_bulk_parity}.

Let us now consider the existence of a hybrid surface-bulk state described by the linear combination
\bea
\Psi_{n,\epsilon}^{\left(\sigma,\sigma'\right)}(\zt)=\alpha\Psi^{(\sigma)}_\text{surface}(\zt,\epsilon)+\beta\phi_\text{bulk}^{(n,\sigma')}(\zt).
\label{eq:ketgeneral}
\eea

\begin{figure}
    \centering
    \includegraphics[width=0.4\textwidth]{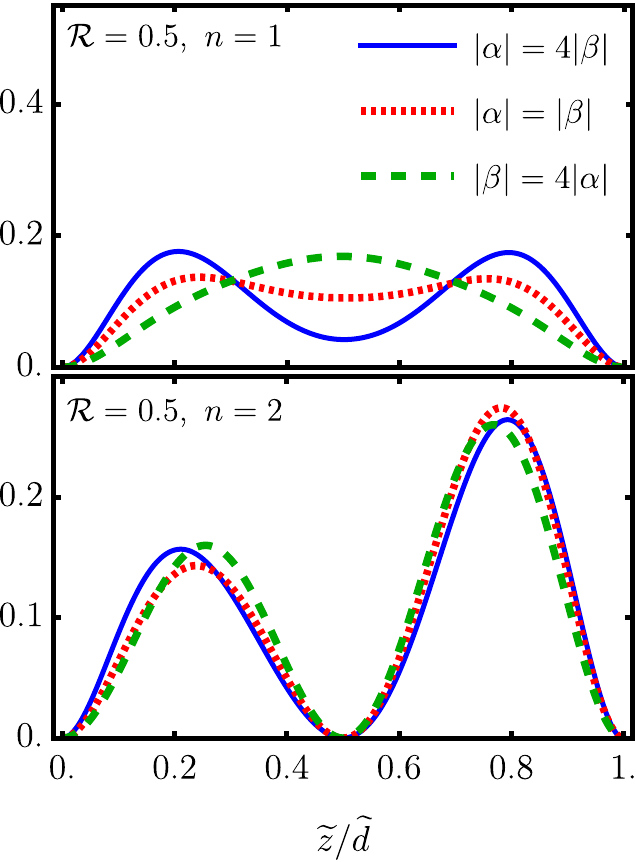}
    \caption{{Spatial distribution of $\rho_{\rm{n,\epsilon}}^{(\uparrow,\uparrow)}(\tilde{z})$ for different coefficients $\alpha$ and $\beta$ in the hybrid state Eq.~\eqref{eq:ketgeneral}.}}
    \label{fig:Psi_updpwn_mix_4}
\end{figure}

The probability distribution corresponding to this hybrid state is given by the expression 
\bea
\rho^{(\sigma,\sigma')}_{n,\epsilon}(z)=\rho_\text{surface}^{(\sigma)}(z)+\rho_\text{bulk}^{(\sigma')}(z)+\rho_\text{mix}^{(\sigma,\sigma')}(z),
\label{eq_rhohybrid}
\eea
where
\bea
\rho_\text{mix}^{(\sigma,\sigma')} = 2\text{Re}\left\{\alpha^*\beta\braket{\text{surface},\sigma|\text{bulk},\sigma'}\right\}
\eea
represents the contribution due to the quantum interference effect between surface and bulk in the hybrid state in Eq.~\eqref{eq:ketgeneral}. As a trivial consequence of Eq.~\eqref{eq_surface_parity} and Eq.~\eqref{eq_bulk_parity}, the probability density contributions  $\rho_{\text{surface}}^{(\sigma)}(z) = |\alpha|^2|\Psi_{\text{surface}}^{(\sigma)}(z,\epsilon)|^2$ and $\rho_{\text{bulk}}^{(\sigma)}(z) =|\beta|^2|\phi_{\text{bulk}}^{(n,\sigma)}(z)|^2$ arising independently from surface and bulk, respectively, are symmetric with respect to the center of the slab, regardless of their spin or parity. This feature is clearly illustrated in the central row of the panel Fig.~\ref{fig:Psi2VsnR1mix}, by the solid (bulk) and dashed (surface) lines, respectively. 

Clearly, for opposite spins $\sigma\ne\sigma'$ the components of the surface and bulk spinor states are mutually orthogonal such that  $\rho_\text{mix}^{(\sigma\ne \sigma')} = 0$, and hence the probability density is indeed trivially symmetric with respect to a mirror plane at the center of the slab $z=d/2$, i.e.
\be
\rho_{n,\epsilon}^{(\sigma\ne\sigma')}(d/2 - \zeta) = \rho_{n,\epsilon}^{(\sigma\ne\sigma')}(d/2 + \zeta),
\ee
for $\zeta \in[-d/2,d/2]$ the distance from the center. This is clearly shown in the first row of the panel Fig.~\ref{fig:Psi2VsnR1mix}, where the dotted line represents the total probability density $\rho^{(\uparrow,\downarrow)}$ for opposite spin components in the hybrid state when $|\alpha| = |\beta|$ in the superposition Eq.~\eqref{eq:ketgeneral}.

{In contrast, a very non-trivial behavior is observed when surface and bulk in the hybrid state Eq.~\eqref{eq:ketgeneral} possess  parallel spins ($\sigma = \sigma'$). In this last case, the mixture term does not vanish, as illustrated in the bottom row of the panel Fig.~\ref{fig:Psi2VsnR1mix}, for the choice $|\alpha| = |\beta|$ in the combination of the hybrid state in Eq.~\eqref{eq:ketgeneral}. Interestingly, given that the surface spinor components possess alternate parities (see Eq.~\eqref{eq_surface_parity}), while the bulk spinor possesses a well defined global parity (see Eq.~\eqref{eq_bulk_parity}), the mixture term $\rho_{{\rm{mix}}}^{(\sigma,\sigma)}$ does not exhibit, in general, a well defined parity for arbitrary values of $n$ and $\mathcal{R}$. However, as can be appreciated in Fig.~\ref{fig:Parity_components} for $\sigma = \uparrow$ and Fig.~\ref{fig:Parity_components_down} for $\sigma = \downarrow$, respectively, when $\R\lesssim 1$ we have that
\bea
\left|\text{Re}\Psi^{(\text{surface},\sigma)}_{j}\right|\ll\left|\text{Re}\Psi^{(\text{surface},\sigma)}_{j+2}\right|,
\eea
and in such case the mix term acquires an approximate symmetry imposed by the parity of the bulk state. This property of the mixture density term is appreciated in Fig.~\ref{fig:Psi_mix2}, for $\R = 0.1$ and different values of $n$, the later determining the parity of the bulk state according to Eq.~\eqref{eq_bulk_parity}.
In consequence, the mirror symmetry of the total probability density
for the hybrid surface-bulk state defined by Eq.~\eqref{eq_rhohybrid} may be broken for parallel spin components, depending on the combination of values of $\R$ and $n$. These occurrences are material-specific because $\R$ depends on the Compton decay length. Consequently, the subtleties of symmetry breaking remain distinctive for each material due to their unique characteristics. A graphical illustration of this effect is observed in the first row of the panel Fig.~\ref{fig:Psi2VsnR1mix}, where the solid line represents the total probability density $\rho^{(\uparrow,\uparrow)}$ for parallel spin components in the hybrid state when $|\alpha| = |\beta|$ in the superposition Eq.~\eqref{eq:ketgeneral}.

This rather counter-intuitive result, given the symmetric boundary conditions in the slab geometry, is clearly a pure quantum mechanical effect due to interference contained in the mixture term.

Figure~\ref{fig:Psi_updpwn_mix_4} highlights the effect of the coefficients $\alpha$ and $\beta$ in the linear combination described by Eq.~\eqref{eq:ketgeneral}. The general considerations regarding inversion symmetry remain unaffected. As expected, larger values of $|\alpha| > |\beta|$ imply an increasingly important contribution from the surface state in the hybrid combination Eq.~\eqref{eq:ketgeneral}, and correspondingly the weight of the probability distribution is more concentrated towards the edges. Conversely, when $|\beta| > |\alpha|$ the probability density of the hybrid state displays more weight towards the bulk. A similar effect is observed as a function of $n$,
since for $n \gg 1$, the hybrid state displays more statistical weight in the bulk.}

\begin{figure}
    \centering
    \includegraphics[scale=0.7]{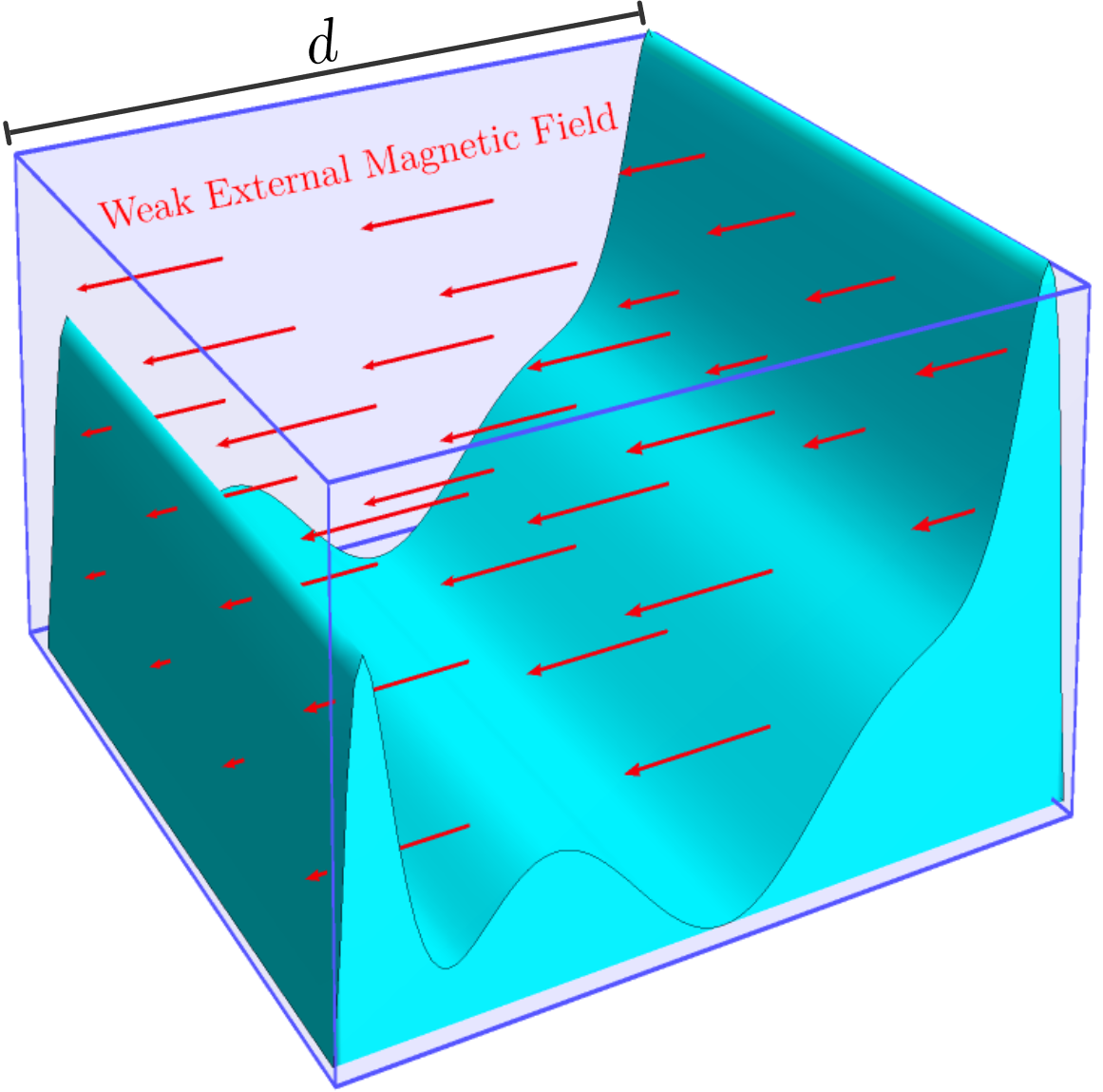}
    \caption{The geometrical configuration discussed in the model: A topological insulator (T.I.) slab of thickness $d$, subjected to a constant and weak magnetic field aligns the spins of surface and bulk electrons in the same direction. This alignment may result in an asymmetric probability density distribution, depicted by the cyan surface.}
    \label{fig:scheme}
\end{figure}

\section{Conclusions}

By means of a standard continuum effective model for a TI in a slab geometry, we explicitly obtained the surface and confined bulk states, in order to investigate the potential effects of their hybridization in the probability density, an important physical parameter that determines the electronic transport properties of such materials near the Fermi level. We assumed symmetric boundary conditions that consistently lead to probability density distributions for the bulk and surface states that possess a mirror symmetry with respect to a plane at the center of the slab.
{Surprisingly, our analytical and numerical results reveal that, when the spin components of surface and bulk in the hybrid state are parallel to each other, the symmetry in the probability density is spontaneously broken under certain conditions, depending on the combination of $n$ (the quantum number in the bulk solution) and the material-dependent parameter $\R$. We can explain this effect in simple terms, due to the spatially asymmetric interference pattern arising between surface and bulk, as captured in the $\rho_{{\text{mix}}}^{(\sigma,\sigma)}$ contribution for parallel spin components. 

In this study, we are not delving into the detailed microscopic mechanism leading to the surface-bulk hybridization, which has been argued to be triggered, for instance, by disorder effects~\cite{PhysRevB.90.245418,PhysRevB.85.121103,Mandal_2023,Yang2020,Hikami_1980}, or analyzed by means of a Fano model with a phenomenological coupling constant~\cite{Hsu_014} in previous works. In contrast, we are here focusing on its possible consequences on the resulting probability density distribution. We found that under certain configurations, when the surface and bulk spin polarizations in the hybrid state are parallel, it may display an asymmetric probability density distribution, a theoretical result not previously reported in the literature~\cite{Hsu_014,PhysRevB.90.245418,PhysRevB.85.121103,Mandal_2023,Yang2020,Hikami_1980}, even though the importance of spin polarization for the emergence of the surface state was already recognized by first-principles calculations in slab geometries~\cite{Reid_2020}.

An experimental scenario where this interesting effect could be observed involves the presence of a constant, weak magnetic field $\mathbf{B} = \hat{z}B$ acting upon the topological insulator slab, as depicted in Fig.~\ref{fig:scheme}. In this scenario, the external field acts as a perturbation that does not alter the system's ground state but breaks its spin degeneracy via the Zeeman coupling. This coupling, involving a magnetic moment $\mu$, results in a linear shift in the corresponding energy eigenvalues described by Eq.~\eqref{Eq.Energydef} as $\epsilon\to\epsilon - s\mu B/2$, where $s=-1$ denotes the $(\uparrow,\downarrow)$ state and $s=+1$ represents the $(\uparrow,\uparrow)$ state. Consequently, under this condition the system tends to favor a hybrid quantum state with parallel spin components for surface and bulk, where we anticipate the emergence of this novel phenomenon. Supporting experimental evidence for this conjecture is indeed presented in the literature~\cite{Buchenau_2017}. For instance, a study on magnetotransport in Bi$_2$Se$_3$ nanoplate devices reveals that the proximity of a magnetic substrate to a surface channel disrupts the symmetry between the top and bottom electronic densities, resulting in the decoupling of the quantum Hall effects on each surface~\cite{Buchenau_2017}.}

\section*{Acknowledgements}

JDCY and EM acknowledge financial support from ANID PIA ANILLO ACT/192023. EM also acknowledges ANID FONDECYT No 1230440.

\section*{Conflict of interest}

The authors declare no conflict of interest.

\section*{Data availability statement}
All data generated or analysed during this study are included in this published article. Any additional information is available from the corresponding author on reasonable request.

\section*{Keywords}
Topological insulators, surface states, bulk states, surface-bulk hybridization, symmetry breaking, electronic density.

\bibliography{BibSurfaceStates}

\end{document}